\documentclass[%
reprint,
superscriptaddress,
 amsmath,amssymb,
 aps,
prb,
]{revtex4-1}

\usepackage{graphicx}
\usepackage{epsfig}
\usepackage{graphics}
\usepackage{color}
\usepackage{dcolumn}
\usepackage{bm}
\usepackage{hyperref}
\usepackage{sidecap}
\usepackage{float}

\begin{document}
\title{Magnon bound states vs. anyonic Majorana excitations in the Kitaev honeycomb magnet $\alpha$-RuCl$_3$}

\author{Dirk Wulferding}
\altaffiliation{Contributed equally to this work.}
\email[]{Corresponding author: dirk.wulferding@tu-bs.de}
\affiliation{Institute for Condensed Matter Physics, TU Braunschweig, D-38106 Braunschweig, Germany}
\affiliation{Laboratory for Emerging Nanometrology (LENA), TU Braunschweig, D-38106 Braunschweig, Germany}

\author{Youngsu Choi}
\altaffiliation{Contributed equally to this work.}
\affiliation{Department of Physics, Chung-Ang University, Seoul 156-756, Republic of Korea}

\author{Seung-Hwan Do}
\affiliation{Department of Physics, Chung-Ang University, Seoul 156-756, Republic of Korea}

\author{Chan Hyeon Lee}
\affiliation{Department of Physics, Chung-Ang University, Seoul 156-756, Republic of Korea}

\author{Peter Lemmens}
\affiliation{Institute for Condensed Matter Physics, TU Braunschweig, D-38106 Braunschweig, Germany}
\affiliation{Laboratory for Emerging Nanometrology (LENA), TU Braunschweig, D-38106 Braunschweig, Germany}

\author{Cl\'{e}ment Faugeras}
\affiliation{LNCMI, CNRS, EMFL, Univ. Grenoble Alpes, 38000 Grenoble, France}

\author{Yann Gallais}
\affiliation{Laboratoire Matériaux et Phénomènes Quantiques (UMR 7162 CNRS), Université Paris Diderot - Paris 7, 75205 Paris cedex 13, France}

\author{Kwang-Yong Choi}
\email[]{Corresponding author: kchoi@cau.ac.kr}
\affiliation{Department of Physics, Chung-Ang University, Seoul 156-756, Republic of Korea}

\date{\today}

\begin{abstract}

The pure Kitaev honeycomb model harbors a quantum spin liquid in zero magnetic fields, while applying finite magnetic fields induces a topological spin liquid with non-Abelian anyonic excitations. This latter phase has been much sought after in Kitaev candidate materials, such as $\alpha$-RuCl$_3$. Currently, two competing scenarios exist for the intermediate field phase of this compound ($B=7-10$ T), based on experimental as well as theoretical results: (i) conventional multiparticle magnetic excitations of integer quantum number vs. (ii) Majorana fermionic excitations of possibly non-Abelian nature with a fractional quantum number. To discriminate between these scenarios a detailed investigation of excitations over a wide field-temperature phase diagram is essential. Here we present Raman spectroscopic data revealing low-energy quasiparticles emerging out of a continuum of fractionalized excitations at intermediate fields, which are contrasted by conventional spin-wave excitations. The temperature evolution of these quasiparticles suggests the formation of bound states out of fractionalized excitations.

\end{abstract}

\maketitle

\section{Introduction}

The search for Majorana fermions in solid-state systems has led to the discovery of several promising candidate materials for exchange-frustrated Kitaev quantum spin systems~\cite{jackeli-09, singh-12, modic-14, chun-15, abramchuk-17, kitagawa-18}. One of the closest realizations of a Kitaev honeycomb lattice is $\alpha$-RuCl$_3$~\cite{plumb-14, sears-15}, where the spin Hamiltonian is dominated by Kitaev interaction $K$. Nevertheless, non-Kitaev interactions, such as Heisenberg ($J$) and off-diagonal symmetric exchange terms (``$\Gamma$-term''), as well as stacking faults in $\alpha$-RuCl$_3$ lead to an antiferromagnetically ordered zigzag ground state below $T_N \approx 7$ K~\cite{sears-15}. The exact strengths of these interactions have not been pinpointed, yet a general consensus on the minimal model has emerged about a ferromagnetic $K \sim$ -6 -- -16 meV, as well as $\Gamma \sim$ 1 -- 7 meV, and $J \sim$ -1 -- -2 meV~\cite{do-17, suzuki-18}. Despite these additional magnetic parameters, the Majorana fermionic quasiparticles are well preserved at high energies and elevated temperatures~\cite{rousochatzakis-18}. Indeed, many independent and complementary experimental techniques have been used to probe the emergence of itinerant Majorana fermions and localized gauge fluxes from the fractionalization of spin degrees of freedom~\cite{glamazda-17, do-17, kasahara-18}.

A promising route in understanding Kitaev physics might be the suppression of long-range magnetic order in a magnetic field, with the possibility of generating an Ising topological quantum spin liquid. For $\alpha$-RuCl$_3$ this field is $B_c \sim 6 - 7$ T~\cite{sears-17, wolter-17, baek-17}. Higher magnetic fields lead to a trivial spin polarized state. In the intermediate field range of 7-10 T, the magnetic order melts into a quantum disordered phase, in which the half-integer quantized thermal Hall conductance is reported~\cite{kasahara-18}. This remarkable finding may be taken as evidence for a field-induced topological spin liquid with chiral Majorana edge states and the central charge $q=\nu/2$ (Chern number $\nu=1$). However, it is less clear whether such a chiral spin liquid state can be stabilized in the presence of a relatively large field and non-Kitaev terms in $\alpha$-RuCl$_3$. In the original Kitaev honeycomb model, non-Abelian Majorana excitations are created upon breaking time-reversal symmetry, e.g., through applying a magnetic field~\cite{kitaev-06}. These composite quasiparticles correspond to bound states of localized fluxes and itinerant Majorana fermions~\cite{theveniaut-17}. The composite bound states are of neither fermionic nor bosonic character, but instead they acquire an additional phase in the wavefunction upon interchanging particles, i.e., they follow anyonic statistics~\cite{stern-10}. Although for a Kitaev system bound itinerant Majorana fermions are possible in the presence of perturbations~\cite{theveniaut-17}, it is unclear how stable a topological spin liquid state is in this case. In particular, an open issue is whether the quantized thermal Hall effect is related to a non-Abelian phase featuring anyonic Majorana excitations.

There exists another scenario in which the intermediate phase is simply a partially polarized phase and smoothly connected to the fully polarized phase. Here the transition through $B_c$ involves conventional multiparticle excitations due to anisotropic interactions~\cite{winter-18}. To resolve these opposing scenarios, one needs to clarify the nature of quasiparticle excitations emergent in the intermediate-to-high-field phase. So far, neutron scattering~\cite{banerjee-18}, THz spectroscopy~\cite{wang-17}, and electron spin resonance~\cite{ponomaryov-17} have revealed a significant reconfiguration of the magnetic response through $B_c$. These methods generally probe $\Delta S = \pm 1$ excitations. Therefore, complementary experiments sensitive to also singlet ($\Delta S = 0$) excitations are essential for unraveling new aspects of low-energy properties and for obtaining a complete picture of individual quasiparticles. In this work, we employ Raman spectroscopy capable of sensing single- and multiparticle excitations over the sufficiently wide ranges of temperatures $T=2-300$ K, fields $B=0-29$ T, and energies $\hbar\omega=1-25$~meV ($8-200$~cm$^{-1}$). At low fields ($B < B_c$) and low temperatures ($T < T_N$) we observe a number of spin wave excitations superimposed onto a continuum of fractionalized excitations. Towards higher fields above 10 T, the magnetic continuum opens progressively a gap and its spectral weight is transferred to well-defined sharp excitations that correspond to one-magnon and magnon bound states, marking the crossover to a field-polarized phase. In the intermediate phase, a weakly bound state emerges. This bound state is formed via a spectral transfer from the fractionalized continuum through an isosbestic point around 8.75 meV, and does not smoothly connect to the magnon bound states in the high-field phase. Our results suggest that this weakly bound state carries Majorana characteristics~\cite{rousochatzakis-18, takikawa-19} and that the intermediate-field phase of $\alpha$-RuCl$_3$ hosts a novel quantum phase.

\section{Results}

\begin{figure*}
\label{figure1}
\centering
\includegraphics[width=16cm]{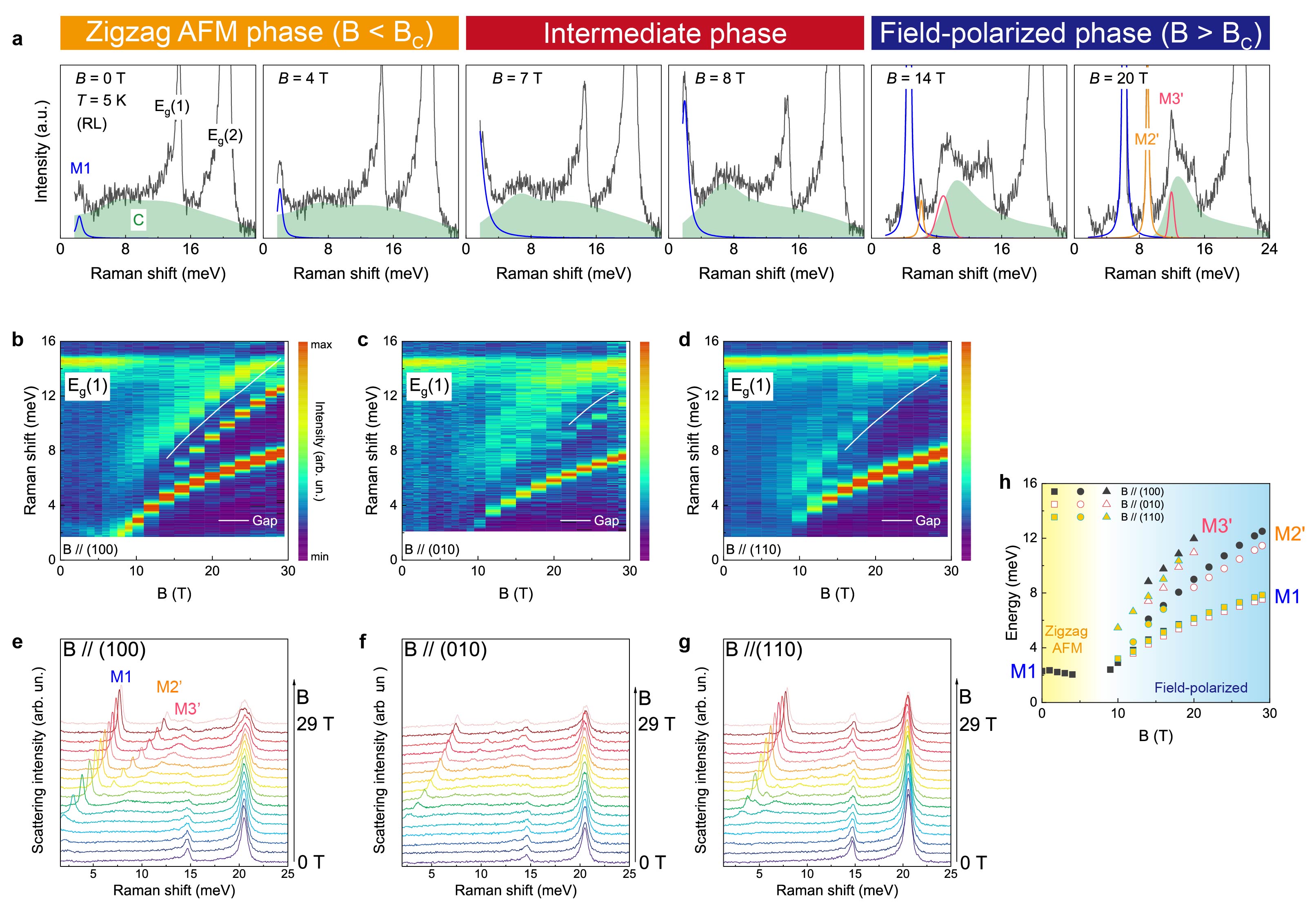}
\caption{\textbf{Field evolution of magnetic excitations in $\alpha$-RuCl$_3$ through field-induced phases.} \textbf{a}, As-measured Raman spectra at $T \approx 5$ K. For $H//a$, $\alpha$-RuCl$_3$ passes successively from a zigzag antiferromagnetic- through an intermediate- to a field-polarized-phase with increasing magnetic fields. The color shading denotes the broad continuum (C) on top of well-defined sharp peaks (M1, M2', M3') and phonon modes ($E_g(1)$ and $E_g(2)$). \textbf{b}, \textbf{c}, and \textbf{d}, Color contour plots of the Raman scattering intensity evolution with magnetic fields aligned along (100), (010), and (110), respectively. \textbf{e}-\textbf{g}, Respective sets of raw Raman data. \textbf{h}, Field-dependence of the sharp low-energy magnetic excitations compared for different field directions.}
\end{figure*}

We performed Raman scattering experiments on oriented single crystals to elucidate the field-evolution of the magnetic excitation spectrum of $\alpha$-RuCl$_3$ (see Supplementary Information, section S1 for a detailed outline of the scattering geometries and section S2 for the full dataset). All measurements were carried out with $RL$ circularly polarized light, probing the $E_g$ symmetry channel. Fig. 1a shows representative raw spectra obtained at increasing fields aligned along the crystallographic $a$ axis [B // (100)]. Besides two sharp, intense phonon modes at 14.5 meV and 20.5 meV [marked $E_g$(1) and $E_g$(2)], we observe several magnetic excitations with a pronounced field dependence: At zero field, the magnetic Raman response consists of a broad continuum (C; green shading) and a sharp peak (M1, blue line). The latter M1 excitation at 2.5 meV is assigned to one-magnon scattering arising from a spin flip process by strong spin-orbit coupling and enables us to detect a gap of low-lying excitations at the $\Gamma$-point as a function of field. The M1 mode energy and its field dependence matches well recent THz magneto-optical data, confirming the $\Delta S = \pm 1$ scattering process as its origin. The green-shaded continuum C agrees with observations in several previous Raman scattering studies~\cite{glamazda-17, sandilands-15, glamazda-16} and has been identified as Majorana fermionic excitations stemming from a fractionalization of spin degrees of freedom in the Kitaev honeycomb model~\cite{kitaev-06}. Although we cannot exclude an incoherent multimagnon contribution to the continuum, the thermal evolution of the continuum follows two-fermionic statistics. Such exotic behavior is not expected for bosonic spin wave excitations, but rather supports the notion of Majorana fermions~\cite{nasu-16}. As detailed in Supplementary Information, section S3, a more prominent two-fermionic character is present at $B_c = 6.7$ T (when zigzag order is suppressed) compared to $B = 0$ T. This can be taken as further evidence for the presence of Majorana fermionic excitations in $\alpha$-RuCl$_3$. As the magnetic field increases above 10 T, C becomes gapped and its spectral width narrows down. This leads to a build-up of spectral weight towards the edge of the gap (solid line in Fig. 1b-d).

Noteworthy is that the gapped continuum has a finite intensity even at high fields $B > B_c$, while several sharp and well-defined excitations emerge additionally with increasing fields. The residual spectral weight of C at sufficiently high fields means that the excitation spectrum in this regime is not solely comprised of single- and multiparticle magnons. Indeed, recent numerical calculations of the Kitaev model under applied fields uncovered a wide Kitaev paramagnetic region, reaching far beyond the critical field at finite temperature~\cite{yoshitake-19}. We therefore ascribe the gapped continuum excitations to fractional quasiparticles pertinent to the Kitaev paramagnetic state. The M1 peak is ubiquitous in all measured fields. The excitation M2' (orange line; the notation ``prime'' is used to differentiate distinct modes in the high field regime $B > B_c$) is split off from the M1 peak above 12 T, while the higher-energy M3' excitation (red line) appears at the lower boundary of the gapped continuum above 10-14 T. In previous experimental field-dependent studies on $\alpha$-RuCl$_3$ ranging from inelastic neutron scattering (INS)~\cite{banerjee-18}, to THz absorption~\cite{wang-17}, to ESR~\cite{ponomaryov-17} similar sharp magnetic excitations were reported and interpreted in terms of one-magnon or magnon bound states. In consideration of the narrow spectral form and energy of the corresponding excitations observed in our data, we assign the M2' peak to a two-magnon bound state and the M3' peak to either a multimagnon excitation or a van Hove singularity of the gapped continuum. The evolution of all magnetic excitations as a function of fields up to 29 T is depicted in the color contour plots of Figs. 1b, 1c, and 1d together with the as-measured Raman spectra in Figs. 1e, 1f, and 1g for field directions along (100), (010), and (110), respectively. A slight anisotropy in magnetic excitations as a function of field-direction becomes apparent, which is highlighted in Fig. 1h. In particular, the energy and field ranges of M3' are sensitive to the in-field directions, indicating the presence of non-negligible in-plane anisotropy terms. Our high-field Raman data evidence the existence of multimagnon excitations and a gapped continuum that characterizes the spin dynamics of the partially polarized phase. The base temperature is however restricted to $T \ge 5$ K in this high-field data. As we show below, a richer spectrum emerges at intermediate fields around $B_c$ upon further cooling.

\begin{figure*}
\label{figure2}
\centering
\includegraphics[width=16cm]{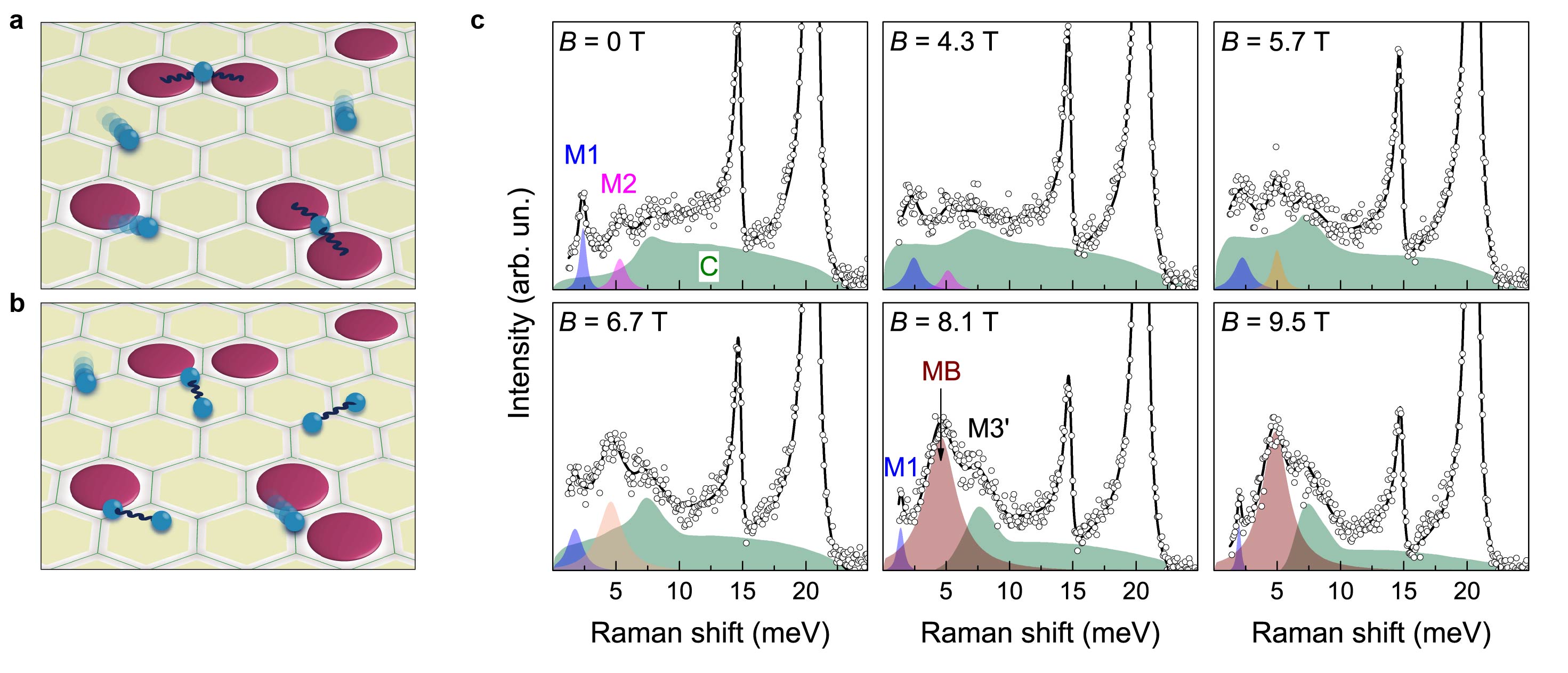}
\caption{\textbf{Magnetic excitations at intermediate magnetic fields.} \textbf{a}, The creation of a bound state from itinerant fermions bound to localized fluxes and \textbf{b} from binding only itinerant Majorana fermions. \textbf{c}, Evolution of Raman data obtained at $T=2$ K (open circles) with increasing fields from 0 T - 9.5 T. The shaded regions denote the decomposition of the magnetic excitations into well-defined peaks and a continuum of excitations. The solid line is a sum of all excitations. The M1 (blue) and M2 (purple shading) modes at low fields of 0 - 4.3 T correspond to spin wave excitations. The excitation MB (dark red) above the critical field of 6.7 T is assigned to a Majorana bound state.}
\end{figure*}

The reported half-integer thermal Hall conductance~\cite{kasahara-18} is expected from chiral Majorana states along the edges of the 2D honeycomb layers of $\alpha$-RuCl$_3$. Simultaneously, anyonic excitations emerge in the bulk. A detection of these anyonic Majorana bound states will provide an ultimate confirmation of the field-induced intermediate non-Abelian phase. In the extended Kitaev system, these bound states can occur in different channels~\cite{theveniaut-17}, namely, from either binding itinerant Majorana fermions to localized fluxes (sketched in Fig. 2a), or by binding two itinerant Majorana fermions (sketched in Fig. 2b). With this in mind, we study the intermediate phase in detail by switching to a magneto-optical cryostat setup, enabling us to reach a base temperature of $T = 2$ K in a field range of $B=0-10$ T. In this setup the sample is tilted by an angle of 18$^{\circ}$ away from the in-plane field geometry, resulting in an additional small but finite out-of-plane field component. Note that here both magnetic field direction as well as light scattering geometry are slightly different from the high-field setup, which prohibits a strict one-to-one comparison of the data. Nonetheless, a good correspondence is found between the high-field B//(110) data at $T \approx 5$ K and the magneto-optical data at $T = 9$ K (see Suppl. Information, section S1). In Fig. 2c we inspect the field-dependence of Raman spectra measured at $T=2$ K. Compared to the $T=5$ K high-field data shown in Fig. 1a, the $T=2$ K data show a new sharp structure at 5 meV (M2) in addition to the one-magnon excitation (M1) and the fractionalized continuum (C). As the field increases, the respective modes evolve in a disparate manner. Initially (at $B=0 - 4.3$ T), the M1 and M2 modes are slightly suppressed, while the continuum C is partially renormalized towards lower energies. As $B_c$ is approached and through 6.7 T, the spectral weight of the continuum is massively redistributed. A new low-energy mode (MB) evolves from the low-field M2 mode with a shoulder structure (M3’) and the continuum of Majorana excitations is gapped above 8.1 T. A recent INS study reported a similarly broad, emerging excitation in the intermediate field-induced phase~\cite{banerjee-18}. It was tentatively discussed as a possible Majorana bound state, but an ultimate assignment was hindered by the lack of a detailed temperature study.

\begin{figure*}
\label{figure3}
\centering
\includegraphics[width=16cm]{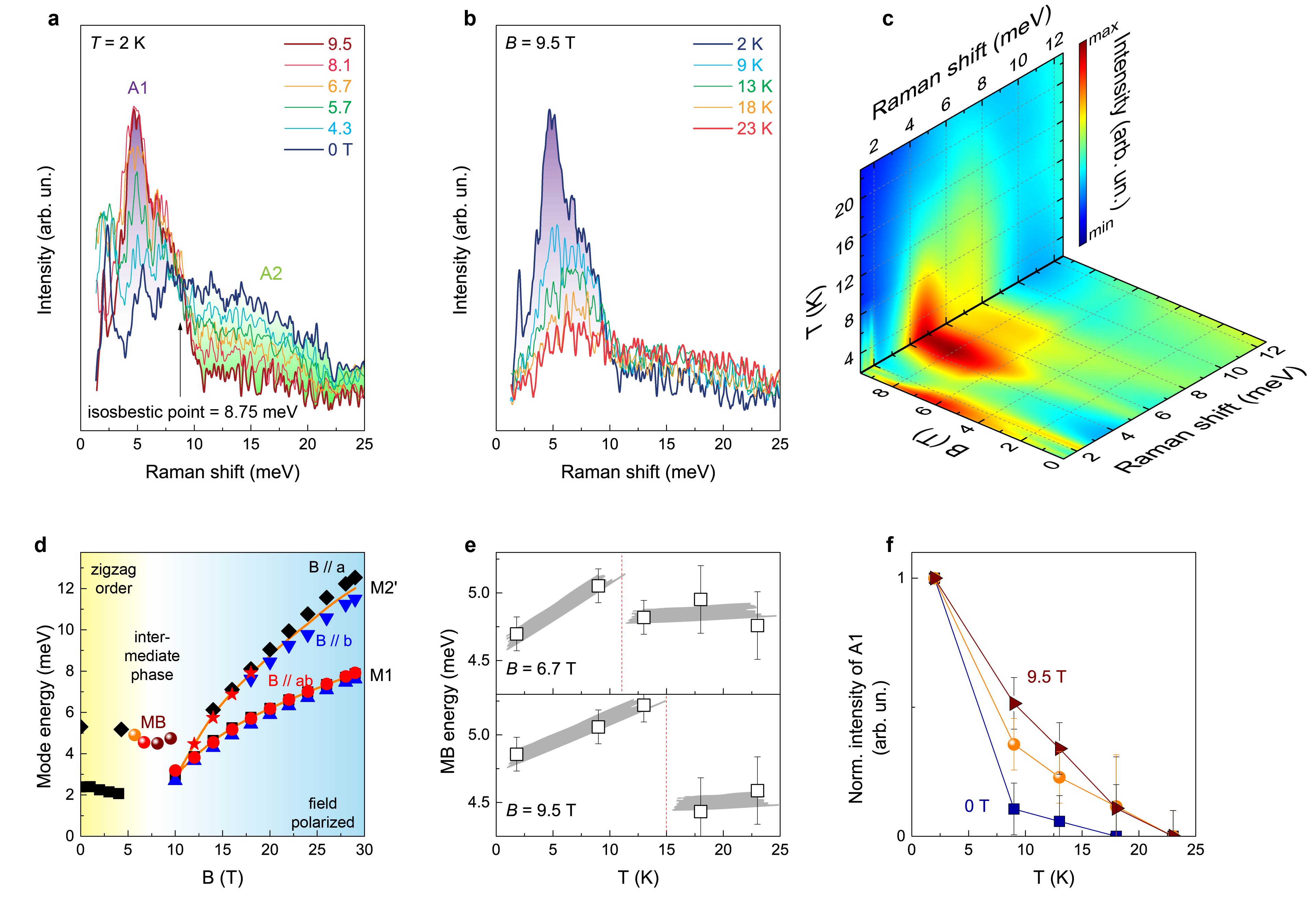}
\caption{\textbf{Spectral weight redistribution and formation of a bound state through $B_c$.} \textbf{a}, Raman spectra obtained at $T=2$ K with increasing magnetic fields. \textbf{b}, Raman spectra at $B=9.5$ T recorded with increasing temperatures. Phonon modes are subtracted in a and b for better visualization of the magnetic excitations. \textbf{c}, Contour plot of the magnetic Raman scattering intensity as a function of temperature and field. \textbf{d}, Field-dependence of excitations around the intermediate phase (spheres) together with spin-wave excitations at low and high magnetic fields. \textbf{e}, Thermally induced binding-unbinding crossover at $B=6.7$ T and 9.5 T. Grey lines are guides to the eyes. \textbf{f}, Thermal melting of the low-temperature magnetic modes at 0 T, 6.7 T, and 9.5 T. Standard deviations in e-f are indicated by error bars.}
\end{figure*}

To analyze the field-induced spectral weight redistribution carefully, we replot phonon-subtracted Raman data taken at $T=2$ K in Fig. 3a. With increasing field a distinct transfer of spectral weight from the mid-energy (green-shaded A2) to the low-energy regime (purple-shaded A1) is observed, with an isosbestic point located around $\omega_{\mathrm{iso}} = 8.75$ meV, at which the magnetic Raman response is independent of the external field. The systematic field-induced redistribution of spectral weight through this isosbestic point suggests an intimate connection between the continuum and the newly formed excitation, and therefore supports the formation of a low-energy Majorana bound state (MB) through a confinement of the high-energy broad continuum of Majorana fermionic excitations. Consistent with the high-field data presented in Fig. 1, we observe a remaining intensity of the continuum C of Majorana fermions at 9.5 T, i.e., corresponding to the non-trivial quantum phase. A coexistence of massive Majorana fermions that form a broad, gapped continuum together with a sub-gap MB state is not compatible with the trivial polarized phase that is characterized by the multimagnon bound states. Figure 3b shows the thermal evolution obtained at $B=9.5$ T (see Suppl. Information, section S4 for full data set). As the temperature increases, the low-energy mode MB gradually loses in intensity and shifts towards higher energies. Meanwhile, the continuum C slightly gains in intensity. The field- and thermal evolution of magnetic modes is visualized in a contour plot in Fig. 3c, based on fits to the as-measured data. It especially highlights the similar evolution of MB with field and temperature: around 5 meV its spectral weight starts to appear at 4.7 T and grows with increasing field, while it becomes thermally stabilized upon cooling below 12 K at 9.5 T. This suggests that quantum and classical fluctuations play the same role in a confinement-deconfinement transition due to a proximate Kitaev paramagnetic state at elevated temperature~\cite{rousochatzakis-18}.

Fig. 3d plots the energy of the MB mode as a function of field together with the spin-wave excitations observed in the zigzag ordered phase as well as in the high field spin-polarized phase. We note a smooth transition from M2 to MB through $B_c$. This weak field-dependence suggests that the MB mode corresponds to an excitation in the singlet sector ($\Delta S_z = 0$), to which Raman spectroscopy is a natural probe. As the magnon corresponds to a condensation of Majorana fermions, the M2-to-MB mode evolution may be interpreted in terms of a condensation-to-confinement crossover where the multimagnon excitations observed at low field gradually evolve into Majorana bound states. Since the continuum C of deconfined Majorana excitations above $B_c$ is massively gapped with an onset energy of 4-6 meV (see Figs. 1b-1d, Fig. 2c), the low-energy bound state can be naturally created within the gap due to confinement. Unlike the M2-to-MB mode crossover below $B_c$, the MB mode is not smoothly linked to the M2' bound state for fields above $B_c$. Rather, as the field increases, the M2' mode splits from the M1 mode, and both excitations are observed prominently, while a signature of the MB mode remains absent in the data obtained from the high-field setup. This suggests that the MB mode is of different nature than the excitations M1 and M2', and that the parameters temperature, scattering geometry, and magnetic field direction play a crucial role in the creation and observation of Majorana bound states. Our data is also contrasted by the rather smooth transition of quasiparticle excitations observed in ESR experiments through 10 T~\cite{ponomaryov-17}, due to the different selection rules for quasiparticle excitations in Raman vs. ESR. The discontinuous evolution observed through 10 T in our data admittedly may be expected in our data due to the slightly different experimental conditions between the high-field and the magneto-cryostat setups, the small temperature difference of $\approx 3$ K cannot account for a jump of 1 meV. We also recall that the vanishing thermal Hall conductance around 10 T parallels the disappearance of the MB mode~\cite{kasahara-18}.

Further support of the MB state interpretation comes from the temperature dependence at two different magnetic fields, 6.7 T and 9.5 T (Fig. 3e). We see a clear initial increase in the MB mode energy at both fields as the temperature rises. This is contrasted by conventional (magnon) unbound excitations, which continuously soften with increasing temperature. For bound states, however, the thermal energy competes with the binding energy~\cite{choi-13}, until eventually an unbinding takes place. Therefore, a sudden drop in mode energy occurs at around 15 K (at $B=9.5$ T), setting the binding energy scale to $\approx 1.3$ meV (see Suppl. Information, section S5 for details). Correspondingly, in a smaller magnetic field of only 6.7 T an unbinding already occurs around $T \approx 10$ K. In contrast, the shoulder of built-up Majorana fermionic excitations gives no clear sign of binding. The thermal evolution of area A1, summarized in Fig. 3f, highlights the gradual melting of the bound state at 9.5 T (dark red triangles) with increasing temperature, while the conventional magnetic excitations at 0 T (blue squares) abruptly vanish above $T_N$. All these observations are consistent with the picture of a quasi-bound state of Majorana fermions that is pulled below the gapped fractionalized continuum by residual interactions of non-Kitaev origin~\cite{knolle-15}.

\section{Discussion}

In the Kitaev model, Majorana bound states are created through flux pairs combined with Majorana fermions in a $\Delta S_z = \pm 1$ channel~\cite{kitaev-06}. However, as flux excitations are largely invisible to the Raman scattering process~\cite{knolle-14}, the bound states between the flux and Majorana fermions barely contribute to the magnetic Raman signal. In the presence of additional non-Kitaev terms, the creation of bound states from itinerant Majorana fermions is enhanced (see the cartoon in Fig. 2b)~\cite{theveniaut-17}. As Raman scattering probes mainly the $\Delta S_z = 0$ channel, we conclude that the MB mode largely consists of the latter Majorana singlet bound states. This interpretation is supported by the smooth crossover from the multimagnon M2 mode to the bound state MB through $B_c$ (see Figure 3d). In such a case, $\alpha$-RuCl$_3$ as an apt realization of the $K-\Gamma$ model (see Suppl. Information, section S5) can host an exotic intermediate-field phase. In relation to this issue, we mention that a numerical study of the $J-K-\Gamma-\Gamma'$ model shows an extended regime of a chiral spin liquid for the out-of-plane field. Once the magnetic field is tilted significantly towards the in-plane direction, the intermediate topological phase vanishes~\cite{gordon-19}. This discrepancy raises the challenging question whether the recently reported field-induced phase has a non-Abelian nature and how the in-plane intermediate phase transits to the alleged chiral spin-liquid phase, if the intermediate phase is of topologically trivial nature.

Our finding demonstrates that a non-trivial crossover from the zigzag through the intermediate to the high-field phase involves a strong reconfiguration of the fractionalized continuum excitations, calling for future work to shed light on the relation between the observed Majorana bound states in an in-plane intermediate field phase of $\alpha$-RuCl$_3$ and the non-Abelian phase predicted for out-of-field directions.

\section{Methods}

\textbf{Crystal growth}
Single crystals of $\alpha$-RuCl$_3$ were grown via a vacuum sublimation method, as described elsewhere~\cite{glamazda-17}. For Raman experiments, samples with dimensions of about $5 \times 2 \times 0.5$ mm$^3$ were selected from the same batch whose thermodynamic and spectroscopic properties have been thoroughly characterized~\cite{glamazda-17, do-17, wolter-17, baek-17}.

\textbf{Raman scattering}
High magnetic fields up to 29 T were generated using the resistive magnet M10 at the LNCMI Grenoble. The sample was kept at a temperature $T \approx 5 - 10$ K and illuminated with a 515 nm solid state laser (ALS Azur Light Systems) at a laser power $P = 0.05$ mW and a spot size of 3 $\mu$m diameter. Resulting Raman spectra were collected in Voigt geometry for in-plane fields, and in Faraday geometry for out-of-plane geometry, using volume Bragg filters (OptiGrate) in transmission geometry and a 70 cm focal distance Princeton Instruments spectrometer equipped with a liquid N$_2$ cooled Pylon CCD camera.

Temperature-dependent Raman scattering experiments in intermediate fields of $B=0-9.5$ T were carried out in 90$^{\circ}$ scattering geometry using a Horiba T64000 triple spectrometer equipped with a Dilor Spectrum One CCD and a Nd:YAG laser emitting at $\lambda = 532$ nm (Torus, Laser Quantum). A $\lambda /4$-plate was used to generate left- and right-circularly polarized light ($RL$). The laser power was kept to $P = 4$ mW with a spot diameter of about 100 $\mu$m to minimize heating effects. A base temperature of $T_{\mathrm{base}} = 2$ K was achieved by fully immersing the sample in superfluid He. Measurements at elevated temperatures were carried out in He gas atmosphere. From a comparison between Stokes- and anti-Stokes scattering we estimate the laser heating to be of 3 K within the He gas environment. The sample temperatures are corrected accordingly. In-plane magnetic fields were applied via an Oxford Spectromag split coil system ($T$ = 2 K -- 300 K, $B_{\mathrm{max}} = 10$ T).

\textbf{Data analysis}
The mid-energy regime of the continuum of Majorana fermionic excitations observed in Raman spectroscopy arises from the simultaneous creation or annihilation of a pair of Majorana fermions. Its temperature dependence can be described by two-fermionic statistics, $I_{\mathrm{MF}} = [1-f(\epsilon_1)][1-f(\epsilon_2)]\delta(\omega-\epsilon_1-\epsilon_2)$; with $f(\epsilon)=1/[1+\mathrm{e}^{\epsilon/k_B T}]$ (see \cite{nasu-16} for details). Additional terms that stem from deviations of the pure Kitaev model ($\Gamma$-term, Heisenberg exchange coupling) culminate in an additional bosonic background term, $I_{\mathrm{B}} = 1/[\mathrm{e}^{\epsilon/k_B T}-1]$. The thermal evolution of the continuum has been fitted to a superposition of both contributions.

Fits to the phonon spectrum were applied using symmetric Lorentzian lineshapes, as well as asymmetric Fano lineshapes~\cite{fano-61} [$I(\omega) = I_0 \frac{(q+\epsilon)^2}{(1+\epsilon^2)}$, with $\epsilon = (\omega-\omega_0)/\Gamma$, and $\Gamma$ = full width at half maximum] in case of a strong coupling between lattice- and spin degrees of freedom. The parameter $1/|q|$ characterizes the degree of asymmetry and - consequently - gives a measure of the coupling strength.

\begin{acknowledgments}
We acknowledge important discussions with Natalia Perkins. Part of this work was performed at the LNCMI, a member of the European Magnetic Field Laboratory (EMFL). This work was supported by ``Nieders\"{a}chsisches Vorab'' through the ``Quantum- and Nano-Metrology (QUANOMET)'' initiative within the project NL-4, DFG-Le967-16, and the Excellence Cluster DFG-EXC 2123 Quantum Frontiers. The work at CAU was supported by the National Research Foundation (NRF) of Korea (Grant No. 2017R1A2B3012642).
\end{acknowledgments}

\section{Supplementary Information}

\section{S1. Scattering geometries}

\begin{figure*}
\label{figure1}
\centering
\includegraphics[width=16cm]{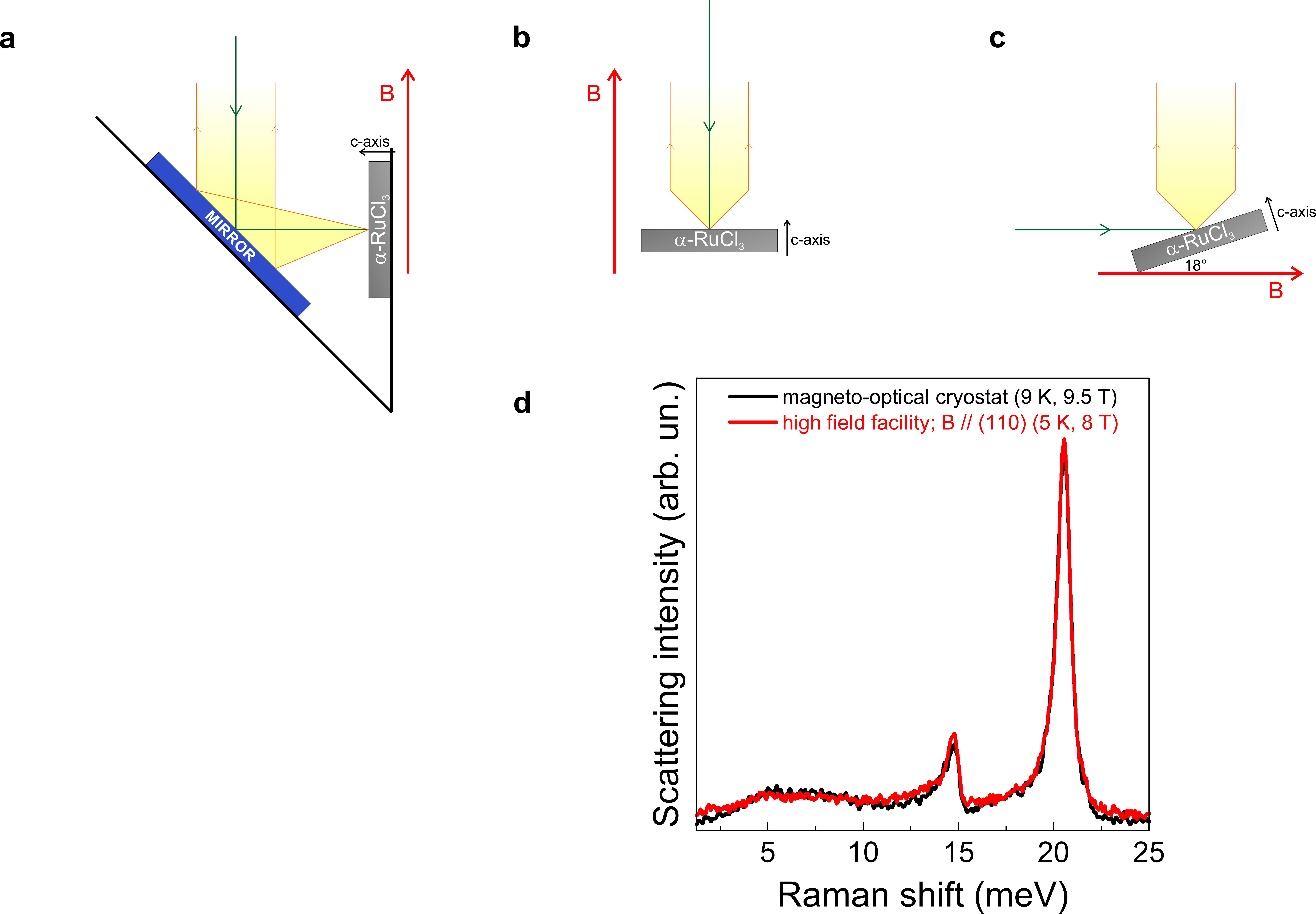}
\caption{\textbf{S1: Sketch of the scattering geometries and data comparison.} \textbf{a}, Raman scattering in Voigt geometry. \textbf{b}, Raman scattering in Faraday geometry. \textbf{c}, 90$^{\circ}$ scattering geometry with the sample tilted by 18$^{\circ}$ from the in-plane field direction for the magneto-cryostat measurements. \textbf{d}, Comparison of Raman data obtained in the magneto-optical cryostat (black line; scattering geometry c) and in the high-field facility (red line, scattering geometry a).}
\end{figure*}

In-plane and out-of-plane Raman scattering experiments at the high magnetic field lab in Grenoble have been carried out in Voigt- and in Faraday geometry, respectively, as sketched in Figs. S1a and S1b. At present, the high-field Raman setup is not equipped with variable temperature inserts (VTIs). For a temperature scan, a magneto-optical cryostat with built-in VTI was used. These experiments were performed in 90$^{\circ}$ scattering geometry (Fig. S1c) in applied magnetic fields up to 10 T. To achieve large enough in-plane field components, the sample was irradiated under a grazing angle of 18$^{\circ}$. A maximum field of 10 T hence corresponds to an in-plane component of 9.5 T. In the main text and in the Supplementary all applied fields have been corrected accordingly. All Raman experiments were performed using circular polarized light: incident right-circular polarized laser light ($R$) was created via a $\lambda$/4 waveplate, and backscattered light was passed through a left-circular polarized analyzer ($L$). This $RL$ configuration allows us to probe the $E_g$ symmetry channel, which carries the magnetic Raman contribution in the pure Kitaev model~\cite{knolle-14}. In Fig. S1d we compare two spectra obtained from the magneto-optical setup and from the high-field facility. Despite the difference in temperature and scattering geometry, we find a nearly one-to-one correspondence.

\section{S2. High field data}

\begin{figure*}
\label{figure2}
\centering
\includegraphics[width=12cm]{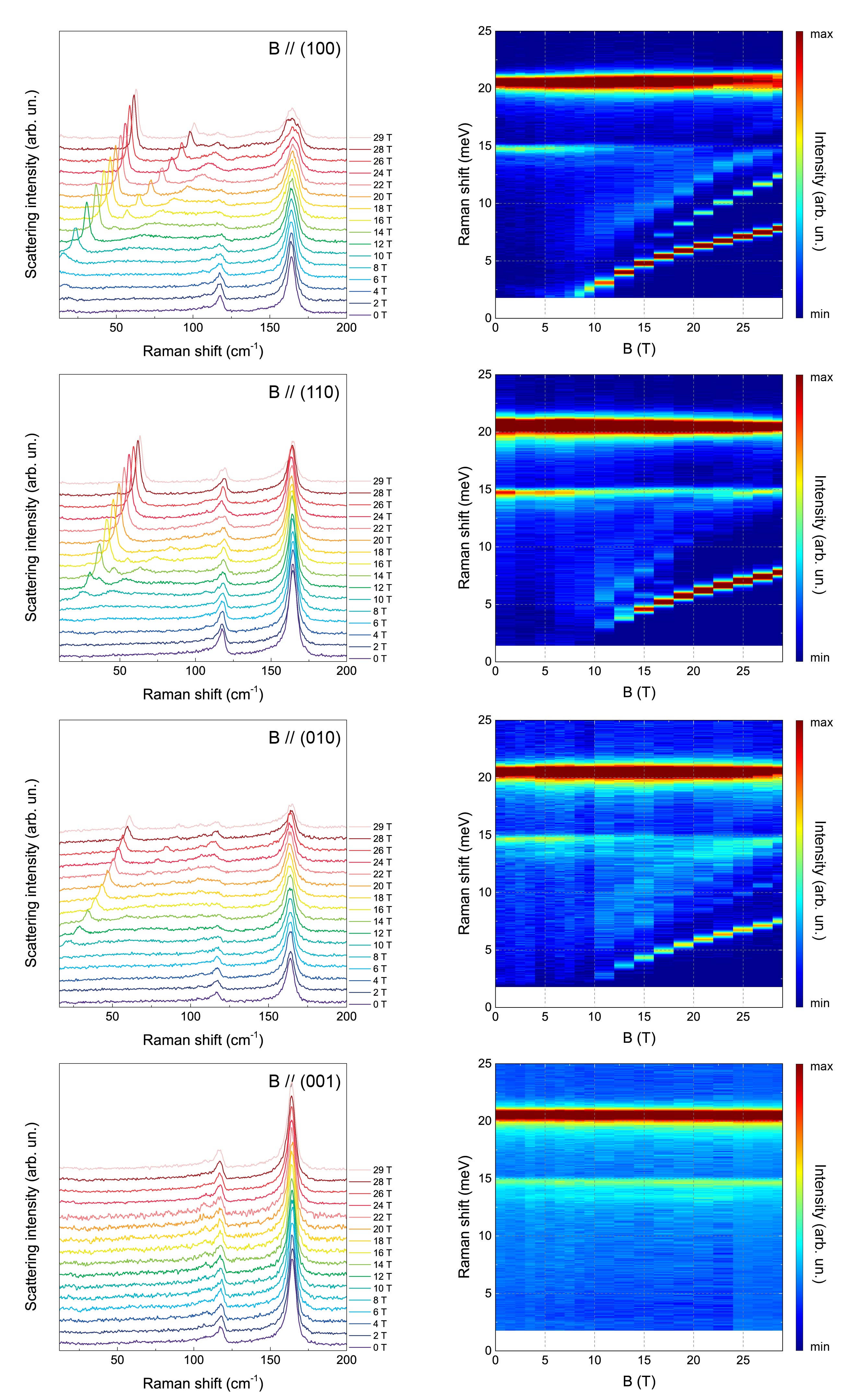}
\caption{\textbf{S2: Raman data obtained over a wide field range in various field directions.} Left panel: As-measured Raman spectra for magnetic fields applied along the (100), (110), (010), and (001) directions. The spectra have been shifted vertically for clarity. Right panel: The respective color contour plots extending up to $E = 25$ meV.}
\end{figure*}

In the left panel of Fig. S2 high-field Raman spectra taken at $T \ge T_N$ are presented for magnetic fields applied along various crystallographic directions. In the right panel the corresponding color contour maps are given. While in-plane magnetic fields along the (100), (010), and (110) directions yield phenomenologically similar behavior concerning the intensity and spectral distribution of spectral weight, the energy scales of the emerging magnetic excitations are slightly different. This is owed to the finite in-plane anisotropy of $\alpha$-RuCl$_3$. 

A substantial spin-phonon-coupling is observed, as the $E_g(1)$ mode gets heavily damped with increasing field through 10 T, while the $E_g(2)$ mode is gradually suppressed. This is owed to the spectral redistribution of C: With increasing magnetic fields the continuum of fractionalized excitations is confined to higher energies as a gap opens and increases in size. In contrast, out-of-plane magnetic fields (along the 001-direction) do not affect the Raman data up to at least 29 T. This is consistent with the anisotropy of the extended Kitaev model.

\section{S3. Deconfined magnetic excitations at the critical field}

\begin{figure*}
\label{figure3}
\centering
\includegraphics[width=16cm]{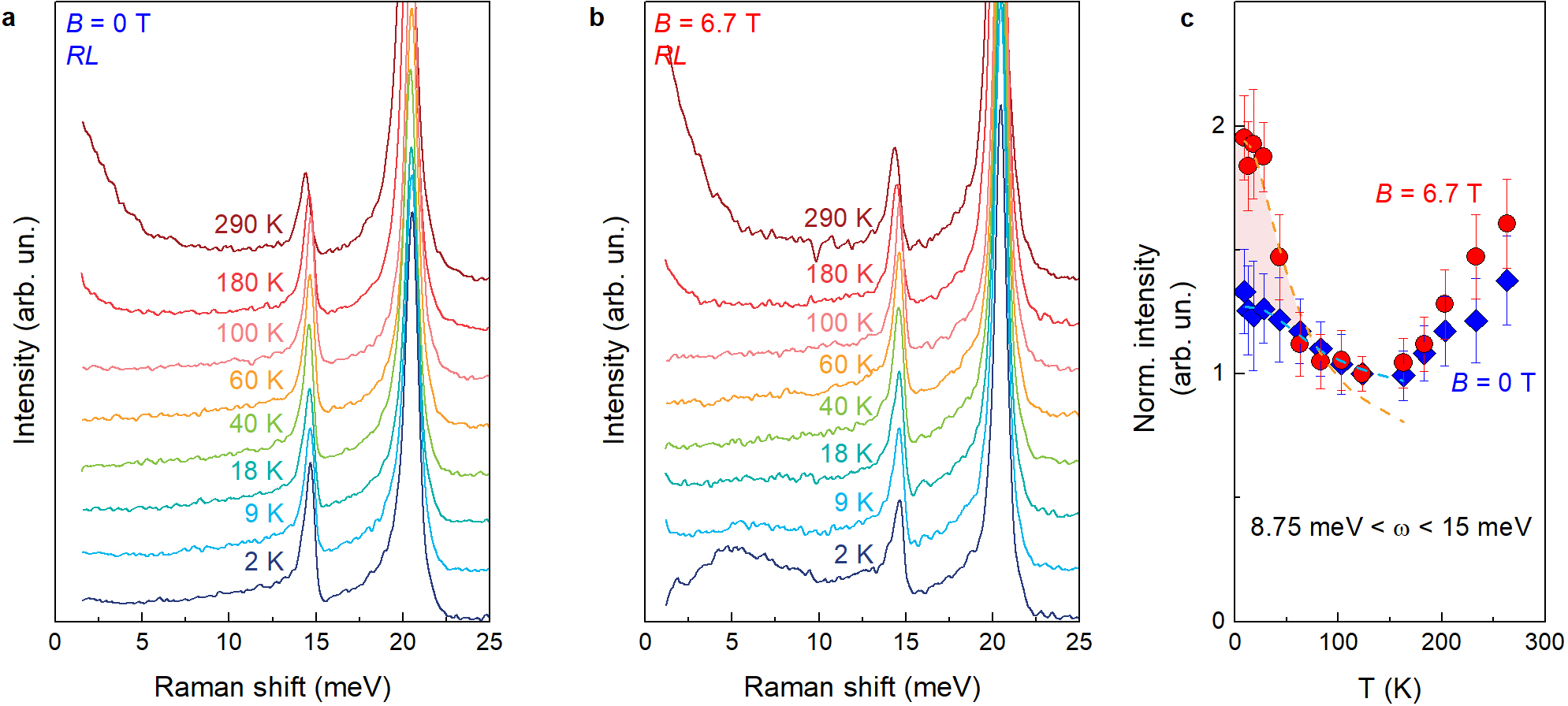}
\caption{\textbf{S3: Temperature dependence and analysis of the quantum statistics.} \textbf{a}, Raman spectra obtained at $B=0$ T in $RL$ polarization over a wide temperature range 2 K -- 290 K. \textbf{b}, Raman spectra obtained around $B_c=6.7$ T in $RL$ polarization over a wide temperature range 2 K -- 290 K. \textbf{c}, Intensity of the broad continuum C integrated over a frequency range of 8.75 -- 15 meV as a function of temperature at $B=0$ T (blue diamonds) and $B=6.7$ T (red circles). Standard deviations are indicated by error bars. The dashed curves represent the two-fermionic contribution.}
\end{figure*}

In order to study the competition of quantum and thermal fluctuations and to analyze the statistics of excitation spectrum C in the intermediate field phase with suppressed LRO we perform experiments as a function of temperature at a fixed magnetic field $B = B_c$. This enables us to distinguish conventional bosonic multiparticle excitations from fractionalized Majorana excitations as quantum fluctuations deconfine the bosonic multiparticle excitations~\cite{nasu-16, sandilands-15, glamazda-16}. In Figs. S3a and S3b we plot Raman spectra obtained at zero fields and around the quantum critical point at $B_c=6.7$ T, respectively, with temperatures ranging from 2 K -- 290 K. The integrated intensity of C from 8.75 meV -- 15 meV is shown in Fig. S3c for $B=0$ (blue diamonds) and $B=B_c$ (red circles). This mid-energy regime of the continuum of Majorana fermionic excitations observed in Raman spectroscopy arises from the simultaneous creation or annihilation of a pair of Majorana fermions. Its temperature dependence can be described by two-fermionic statistics, $I_{\mathrm{MF}} = [1-f(\epsilon_1)][1-f(\epsilon_2)]\delta(\omega-\epsilon_1-\epsilon_2)$; with $f(\epsilon)=1/[1+\mathrm{e}^{\epsilon/k_B T}]$ (see \cite{nasu-16} for details). Additional terms that stem from deviations of the pure Kitaev model ($\Gamma$-term, Heisenberg exchange coupling) culminate in an additional bosonic background term, $I_{\mathrm{B}} = 1/[\mathrm{e}^{\epsilon/k_B T}-1]$. The thermal evolution of the continuum has been fitted to a superposition of both contributions. A more dramatic rise in intensity towards low temperatures is evident in the vicinity of quantum criticality. Similarly, the Fano asymmetry of the $E_g(1)$ phonon overlapping with C is more pronounced at $B_c$, indicative of a stronger coupling between lattice and Majorana fermionic excitations. Our observations suggest that the Majorana-related excitations at the quantum critical field $B_c$ become more pronounced than the $B=0$ T excitation in spite of the development of the low-energy Majorana bound state MB. This suggests that the field-induced phase is thus closer to a spin liquid than the zero-field phase.

\section{S4. Temperature- and field dependence}

\begin{figure*}
\label{figure4}
\centering
\includegraphics[width=16cm]{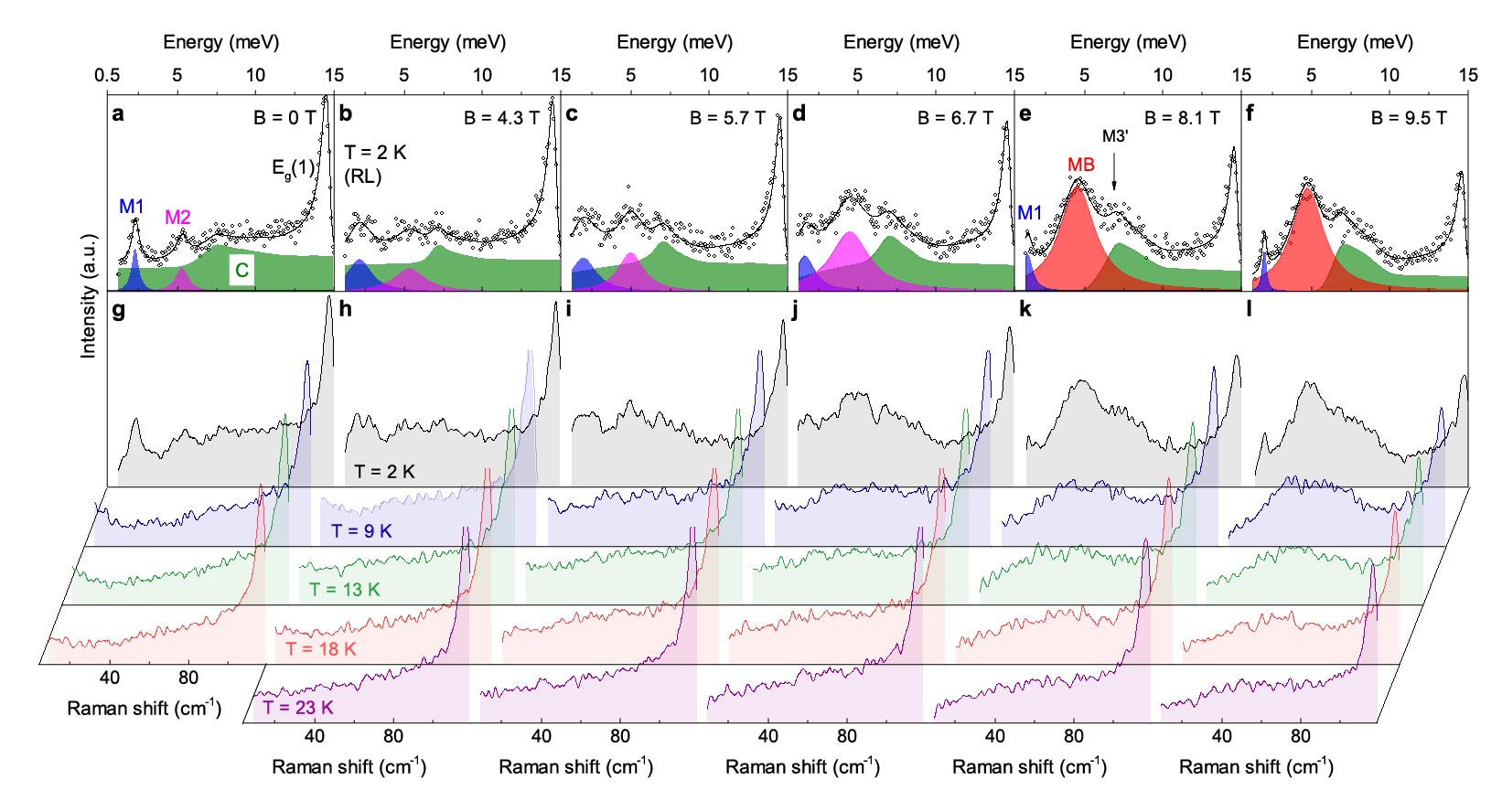}
\caption{\textbf{S4: Field- and temperature dependence of Raman spectra.} \textbf{a}-\textbf{f}, Field dependence of Raman spectra at $T = 2$ K and in circular $RL$ polarization. With increasing external fields through the critical regime $B_c \approx 6.7$ T, the gapless continuum (C; green shading) becomes gapped, while the higher-energy magnon mode (M2; purple) evolves into a bound state (MB). In contrast, the lower-energy magnon mode (M1; blue) persists across $B_c$. \textbf{g}-\textbf{l}, Thermal evolution of Raman spectra at the given applied magnetic fields.}
\end{figure*}

In Fig. S4 we present the full dataset of temperature- and field-dependent as-measured Raman data. The upper panel (a-f) presents the decomposition into various magnetic excitations at base temperature and with increasing magnetic field (see also main text). In Figs. S4g-l the thermal evolution at a given field is shown. For $B < B_c = 6.7$ T, the magnon modes (M1 and M2) rapidly disappear with increasing temperature above $T_N$. For $B > B_c$, the bound state (MB) is gradually suppressed with temperature, while a finite remnant spectral weight about 6.25 meV is observed even at $T=23$ K. This suggests that the nature of low-energy excitations is altered through $B_c$.

\section{S5. $K-\Gamma$ model: Theory vs Experiment}

\begin{figure*}
\label{figure5}
\centering
\includegraphics[width=10cm]{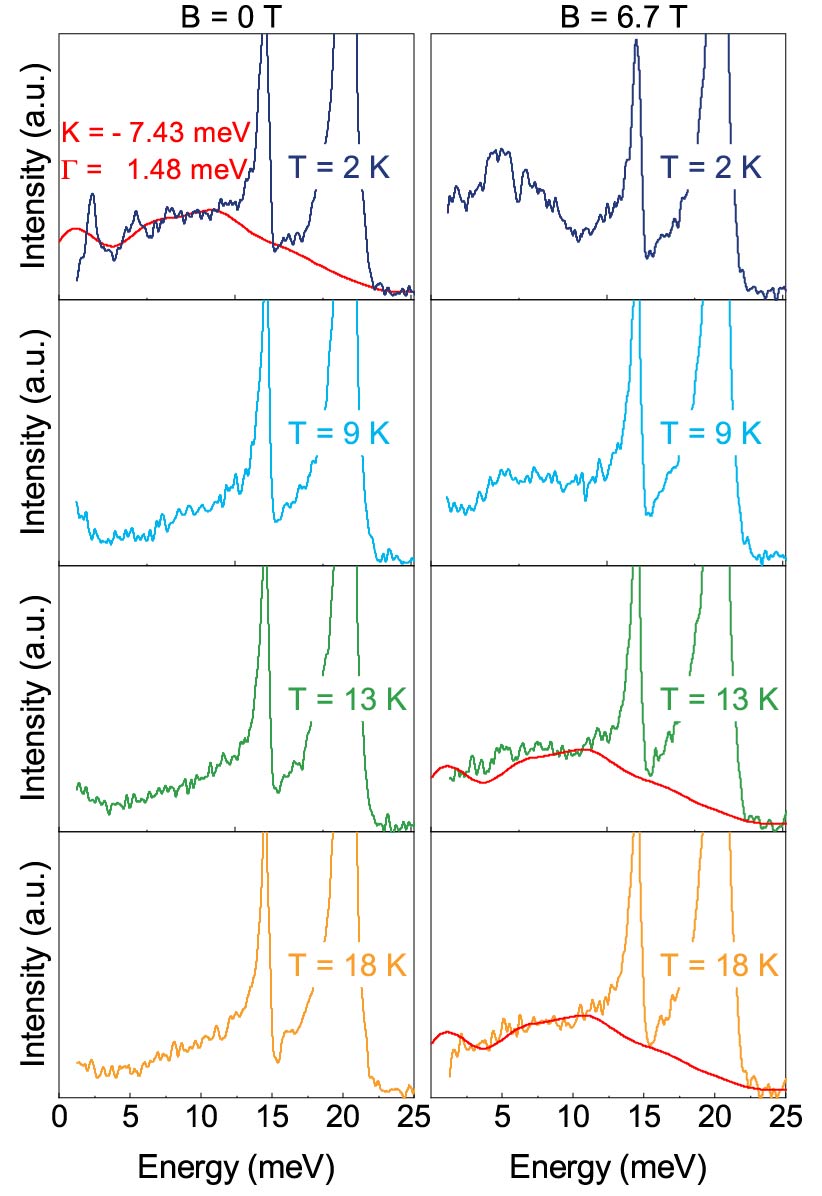}
\caption{\textbf{S5: $K-\Gamma$ Model, Experiment vs. Theory.} As-measured temperature- and field-dependent Raman data together with calculations of the spectral weight based on the $K-\Gamma$ model~\cite{rousochatzakis-18} (solid red line). The parameters to fit the data were chosen as $K = -7.43$ meV and $\Gamma = 1.48$ meV, and the calculations were performed for small but finite temperatures.}
\end{figure*}

Recent theoretical modeling of the Raman response in a Kitaev magnet has been extended to include a $\Gamma$-term. The key impact of the considerable $\Gamma$-term is to shift the spectral weight of itinerant Majorana fermionic excitations to a lower energy~\cite{rousochatzakis-18}. More specifically, the $\Gamma$-term on the one hand generates strong confinement (by opening a massive excitation gap), and on the other hand, rearranges the spectral weight towards lower energies, offering abundant low-energy states. Consequently, the $K-\Gamma$ model is in favor of stabilizing low-energy bound states. At zero fields, the $K-\Gamma$ model gives a reasonable description of our $T=2$ K data (see Fig. S5) as the magnon modes contribute only small spectral weight. In contrast, at $B_c$, the itinerant Majarana fermions are confined to form the bound state at low temperatures. As a consequence, the $K-\Gamma$ model reproduces the Majorana continuum excitation at around the temperature where the Majorana fermions start to bind ($T=13-18$ K). The comparison between our data and the $K-\Gamma$ model theory strongly supports the notion that the $K-\Gamma$ model can capture a key feature of the $\alpha$-RuCl$_3$ magnetism possibly because the residual perturbations are weaker than the $K$ and the $\Gamma$ terms, and can be effectively quenched in applied magnetic fields and at slightly elevated temperature (see Fig. S5).

Based on calculations of the dynamical spin response it was shown that with increasing strength of the anisotropy $\Gamma$-term a second mode can occur as a consequence of an inequality in gap size $\Delta_x$ and $\Delta_z$~\cite{knolle-15}. This might account for the rather large linewidth of MB (see, e.g., Fig. S4e). From an analysis of the $T$ dependence of the peak energy (see Fig. 3e, main text), we estimate the binding energy as $E_\mathrm{B} = \omega_{\mathrm{MB}} - \Delta$, where $\omega_{\mathrm{MB}}=5$ meV and $\Delta$ are the energy of MB and the excitation gap of the continuum C, respectively. With a binding-unbinding transition temperature around 15 K we estimate $E_B$ to be around 1.3 meV, which places the gap energy around $\Delta = 6.25$ meV. This estimation is in very good agreement with the fit of the continuum in Fig. 2c, main text (at $B = 9.5$ T). It furthermore implies that the enhanced spectral weight around 8 meV lies above $\Delta$ and is part of the field-induced spectral re-distribution of Majorana fermionic excitations, indicative of an opening of the large energy gap due to the sizable $\Gamma$-term.

\end{document}